\documentclass[epj]{svjour}
\usepackage{graphics}
\usepackage{graphicx}
\usepackage{color}
\begin{document}
\title{Membership in social networks and the application in information filtering}
\author{Wei Zeng\inst{1,2} \and An Zeng\inst{2}\thanks{\email{an.zeng@unifr.ch}} \and Ming-Sheng Shang\inst{1,3}\thanks{\email{shang.mingsheng@gmail.com}} \and Yi-Cheng Zhang\inst{1,2,3}}
\institute{Web Sciences Center, School of Computer Science and Engineering, University of Electronic Science and Technology of China, Chengdu 611731, China \and Department of Physics, University of Fribourg, Chemin du Mus\'{e}e 3, CH-1700 Fribourg, Switzerland \and Institute of Information Economy, Hangzhou Normal University, Hangzhou 310036, China}

\date{Received: date / Revised version: date}

\abstract{
During the past a few years, users' membership in the online system (i.e. the social groups
that online users joined) are wildly investigated. Most of these works focus on the
detection, formulation and growth of online communities. In this paper, we study users'
membership in a coupled system which contains user-group and user-object bipartite
networks. By linking users' membership information and their object selection,
we find that the users who have collected only a few objects are more likely to be ``influenced"
by the membership when choosing objects. Moreover, we observe that some users may
join many online communities though they collected few objects. Based on these findings,
we design a social diffusion recommendation algorithm which can effectively solve the user
cold-start problem. Finally, we propose a personalized combination of our method and the hybrid method
in [PNAS 107, 4511 (2010)], which leads to a further improvement in the overall recommendation performance.
} 
\maketitle
\section{Introduction}
Clustering is one of the most important features in social systems. Network representation of these systems allows us to identify communities which are distinguished by the density of links higher in communities than among them~\cite{Santo:2010}. In the past, many methods have been proposed to detect these kind of structure-based communities, such as modularity maximizing method~\cite{Newman:2004}, signaling method~\cite{Hu:2008} and Spectral clustering method~\cite{Donetti:2004}. Recently, a significant community structure was detected in the online marketing network \cite{Reichardt:2007}. Communities have concrete applications \cite{Santo:2010}. For instance, one can set up efficient recommendation systems by identifying clusters of customers with similar interests in the customers-products bipartite network (e.g. www.amazon.com) \cite{Reddy:2002}. Moreover, community information and link prediction can benefit from each other~\cite{Yan:2012,Yan:2012II,Zeng:2012PRE}.

Recently, another kind of community in online commercial systems also received much attention. In some online social networking sites, such as MySpace and LiveJournal, users with the same interest can join one group and share information with each other. Due to the rapid development of these sites, this kind of online groups are becoming increasingly prominent and wildly investigated. These explicit user-defined groups are quite different from the structure-based communities. The online groups are created by online individuals while the structure community are detected by algorithms. Yang \cite{Yang:ICDM:2012} defined this kinds of groups as \emph{ground-truth communities}. In ref. \cite{Backstrom:2006:KDD}, the formation of online groups in two large social networks is studied. They show that the tendency of an individual to join a group is influenced not just by the number of friends he/she has within the group, but also crucially affected by how those friends are connected to each other. Besides, ref. \cite{Kairam:2012:WSDM} studies the life and death of online groups. They find that a group attracts new members through the friendship ties of its current members to outsiders.

To the best of our knowledge, few work investigates the relation between the online groups and users' choices of objects so far. In reality, it always happens that a user make his/her decision on choosing an object by referring to other users' comments. In this case, the comments from the group mates can be very valuable to the user. Besides, one can find like-minded users in the online group and directly adopts the items selected by these users. Therefore, the membership (i.e. the social groups that online users joined) will inevitably influence individuals' choices of objects. However, this process is not easy to be directly studied since there is no record about through which way users select objects in real cases. In this paper, we adopt a mathematical framework called the \emph{Aspect Model} which allows us to calculate the potential influence of the social grouping on users' selection of objects \cite{Hofmann:2003:SIGIR}. Specifically, we compute the probability that a user can find his/her interested objects from the group mates' selected objects. If the probability is high, we consider the potential influence of groups on this user's choice of object to be large. The analysis is based on a coupled system which includes both the user-group and user-object bipartite networks. Our results show that small degree users are more likely to be influenced by the membership than large degree users. Moreover, we observe from empirical data that some users who have collected a few objects may join many groups. The results suggest that the data of social groups can be very valuable in information filtering~\cite{Yuan:2011:RS}.

Recommendation, as a typical information filtering problem, has been intensively studied in recent years by physicists~\cite{Lv:RP:2012,Cimini:2011:EPJB,Chen:2013:EPJB,Zhang:JSM:2010,Shang:EPL:2010,Lv:PRE:2011,EPL10058005,PhysicaA3911822,Zhang:EPL:2010}. One of the biggest challenges in recommendation is the user cold-start problem. In online systems, the new/inactive users have only expressed few ratings or collected few objects, which makes recommendation algorithms fail to accurately predict the objects these users interested in. In real online systems, web sites are competing for users. In order to attract more users, web sites should provide new/inactive users with more accurate recommendations. Therefore, addressing the user cold-start is of great importance from the commercial point of view. Based on the user-group and user-object bipartite networks, we propose a social diffusion recommendation algorithm which is able to provide much more accurate recommendations for those small degree users than some well-know methods~\cite{Adomavicius:2005}. Finally, we propose a personalized combination of our method and the hybrid method~\cite{Zhou:PNAS:2010}, which leads to a further improvement in the overall recommendation performance.

\section{Data set and empirical analysis}
\label{Sec:dataset}
In the online system we considered, users not only select objects but also join in groups they are interested in. Such online system can be represented by two bipartite networks $G(U,O,E)$ and $G'(U,C,E')$, where $U=\{u_1, u_2, ..., u_m\}$, $O=\{o_1, o_2, ..., o_n\}$ and $C=\{c_1, c_2, ..., c_l\}$ denote the sets of users, objects and communities, respectively. $E=\{e_1, e_2, ..., e_p\}$ is the set of links between users and objects and $E'=\{e'_1, e'_2, ..., e'_q\}$ is the set of links between users and groups. These two networks can be represented by two adjacency matrixes by $A_{m \times n}$ and $B_{m \times l}$, respectively. The element $a_{i \alpha}$ in $A_{m \times n}$ equals to $1$ if user $i$ collected $\alpha$ and $0$ otherwise, the element $b_{ic}$ in $B_{m \times l}$ equals to 1 if user $i$ joined the group $c$ and 0 otherwise.

Two datasets are investigated here: \emph{Last.fm}\footnote{http://www.last.fm} and \emph{Douban}\footnote{http://www.douban.com}. The Last.fm is a worldwide popular social music site. As mentioned above, the data we used in this paper consists of two types of data: user-object data and user-group data. The object in the dataset is referred to the artist and the membership refers to the online groups users joined. Douban, launched on March 6, 2005, is a Chinese Web 2.0 web site providing user with rating, review and recommendation services for movies, books and music. It is one of the largest online communities in China. Users can view the movies, books and music and also assign 5-scale integer ratings (from 1 to 5) to them. In this paper, we only collect users' activities on the movies and all the groups with movie tag. We treat the user-object interaction matrix as binary, that is, the element equals to 1 if the user has viewed or rated the object and 0 otherwise. In
both systems, we only sample the users who joined at least one group. The basic statistics of these two datasets are presented in the Table \ref{Tab:data_statistics}.

\begin{table*}
\centering \caption{The statistics of datasets.}\label{Tab:data_statistics}
\begin{tabular}{l|c|c|c|c|c}
\hline \hline
     Dataset & \#user & \#objects & \#groups & \#user-objects pairs & \#user-group paris \\[2pt]
    \hline
    Last.fm & 27,500 & 22,443 & 20,341 & 1,503,938 & 309,633 \\[2pt]
    Douban & 25,039 & 25,182 & 2,635 & 2,107,251 & 121,810 \\[2pt]
    \hline\hline
\end{tabular}
\end{table*}

The degree of an object $k_o$ is defined as the number of user who collected it and the degree of a group $k_c$ is defined as the number of users who joined it. For these two datasets, both the object degree and group degree distribution follow the power-law form. As shwon in Table \ref{Tab:data_statistics}, the user-group network is much sparser than the user-object network. In this sense, the information extracted from membership identity is relatively limited. However, we will show in next section that this information is crucial to improve the accuracy of object recommendation, especially for small degree users.

\begin{figure}
\centering
\includegraphics[width=9cm,height=8cm]{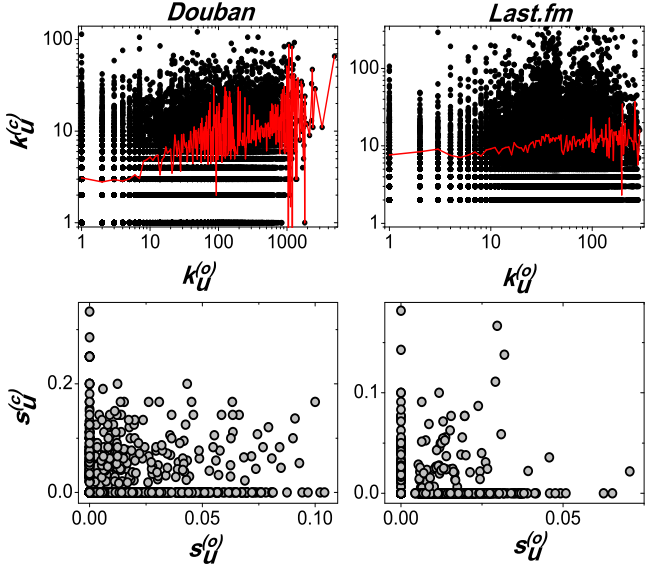}
\caption{The top sub-figures demonstrate the correlation between the number of a user's collected objects $k_u^{(o)}$ and joined groups $k_u^{(c)}$. Given a $x$, the red line is obtained by averaging $k_u^{(c)}$ over all users whose $k_u^{(o)}$ equals $x$. The bottom sub-figures show the correlation between users' similarities based on collected objects $s_{ij}^{(o)}$ and joined groups $s_{ij}^{(c)}$.}
\label{Fig:network_correlation}
\end{figure}

The degree of a user with respect to the objects is denoted as $k^{(o)}_u$ and the degree of a user with respect to the groups is denoted as $k^{(c)}_u$. We present the correlations between $k^{(o)}_u$ and $k^{(c)}_u$ in the top sub-figures of fig. \ref{Fig:network_correlation}. From the scatter plot, it is clear that there are many users who selected few objects but joined many groups. This is confirmed by the averaged curves (see the red curves). Generally speaking, the correlations between $k_u^{(o)}$ and $k_u^{(c)}$ in both datasets is weak. Actually, the weak correlation is due to the property of the real systems. Watching a movie or listening to an album is relatively time consuming. Before selecting a movie to watch or an album to listen, a user will check some background information about these objects to make sure they are sufficiently good. In order to get such information, some users may join many relevant groups and read the group mates' comments. For those users, they may join in many groups but collect only a few objects.

We further compare different users' similarities based on their collected objects and joined groups. Given two users $i$ and $j$, one can calculate their similarity by comparing their collected objects as
\begin{equation}
s^{(o)}_{ij}=\frac{|\Gamma^{(o)}_i \bigcap \Gamma^{(o)}_j|}{|\Gamma^{(o)}_i \bigcup \Gamma^{(o)}_j|},
\label{eq:jaccard_o}
\end{equation}
where $\Gamma_i^{(o)}$ and $\Gamma_j^{(o)}$ denotes the collected object set of $i$ and $j$, respectively. Besides, we can calculate the similarity between users based on their joined groups,
\begin{equation}
s^{(c)}_{ij}=\frac{|\Gamma^{(c)}_i \bigcap \Gamma^{(c)}_j|}{|\Gamma^{(c)}_i \bigcup \Gamma^{(c)}_j|},
\label{eq:jaccard_g}
\end{equation}
where $\Gamma_i^{(c)}$ and $\Gamma_j^{(c)}$ denote the joined group set of $i$ and $j$, respectively. For the sake of clear presentation, we sample $50$ users randomly and report the correlation between $s^{(o)}_{ij}$ and $s^{(c)}_{ij}$ in the bottom sub-figures of fig. \ref{Fig:network_correlation}. Again, there is no strong correlation.

\begin{figure}
\centering
\includegraphics[width=9cm,height=5cm]{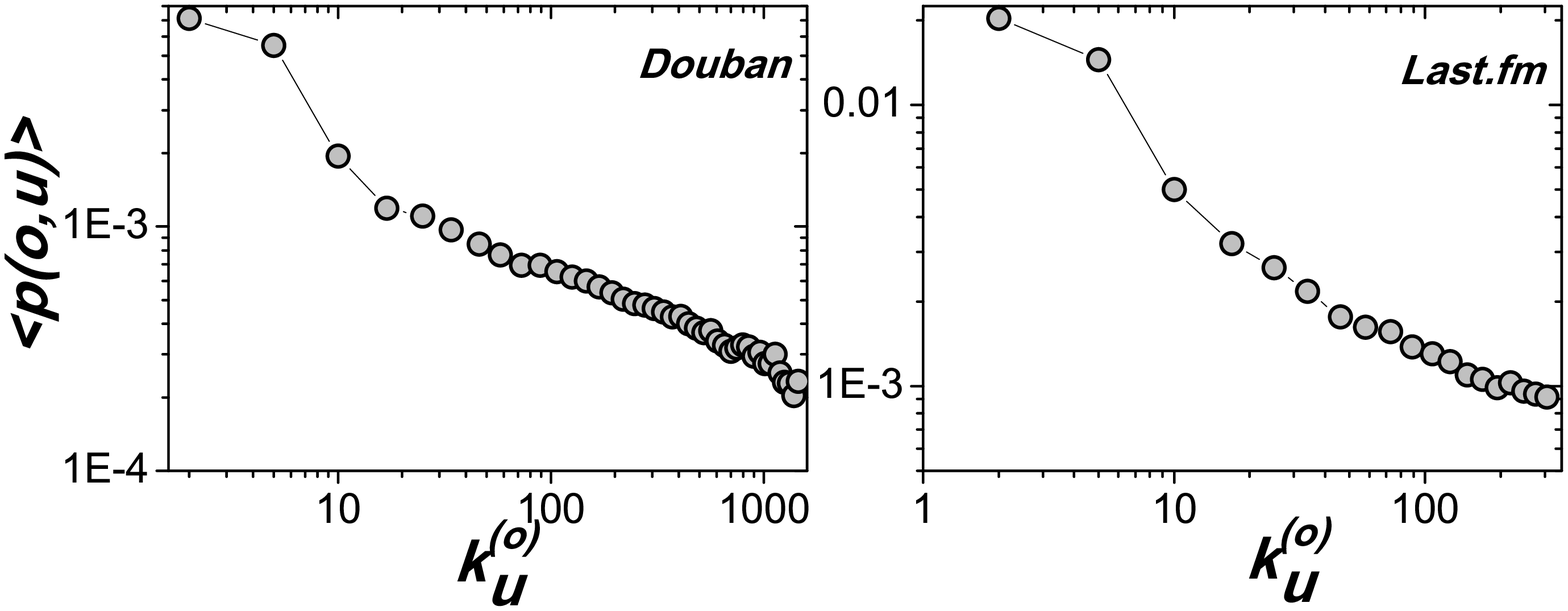}
\caption{The dependence of $\langle p(o,u)\rangle$ on the $k_u^{(o)}$. For a given $x$, its corresponding $\langle p(o,u)\rangle$ is obtained by averaging all the users whose $k_u^{(o)}$ are in the range of $[a(x^2-x), a(x^2+x)]$, where $a$ is chosen as $0.5\log5$.}
\label{Fig:select_rate}
\end{figure}

In next step, we will investigate the potential influence of the social grouping on users' selection of objects. The analysis is based on the model in ref~\cite{Hofmann:2003:SIGIR}. In the model, a user $u$ may participate probabilistically in one or more virtual groups (denoted by $z$) and his/her selection of items is assumed to be linked with these groups. As an example, users in the group $z$ may introduce an object $o$ to user $u$ and $u$ may select or rate it. Therefore, each rating is modeled as a mixture of these virtual groups, which is given by
\begin{equation}
p(o,u) = \sum_z{p(o|z)p(z|u)p(u)},
\label{eq:lsi}
\end{equation}
where $p(u)$ is the probability to select user $u$, $p(z|u)$ is the probability to pick a group $z$ from $u$'s joined groups and $p(o|z)$ is the probability to pick an object $o$ in all the objects selected by users in group $z$. In ref.~\cite{Hofmann:2003:SIGIR}, $p(o,u)$ was claimed to be the potential probability that a user $u$ select an item $o$ through these virtual groups. Actually, $p(o,u)$ can be also regarded as the probability that user $u$ find his/her interested object $o$ from the group mates' selected objects. If the probability is high, the potential influence of groups on this user's choice of object is considered to be large. Since the datasets used in this paper consist explicit user-defined groups, the virtual group $z$ can be naturally replaced by the real group $c$ and the equation \ref{eq:lsi} can be rewritten as
\begin{equation}
p(o,u) = \sum_{c \in C}{p(o|c)p(c|u)p(u)},
\end{equation}
where $C$ is the set of groups that user $u$ has joined, $p(c|u)$ is the probability to pick a group $c$ from $u$'s joined groups and $p(o|c)$ is the probability to pick an object $o$ in all the objects selected by users in group $c$. Suppose the chance of each group to be selected from a user $u$'s joined groups set is equal, $p(c|u)=1/k^{(c)}_u$. $p(o|c)$ is the probability that object $o$ is selected by users in group $c$:
\begin{equation}
p(o|c) = \frac{\sum_{u \in c}{a_{uo}}}{\sum_{u \in c}{a_{u\cdot}}}=\frac{\sum_{u \in c}{a_{uo_{\alpha}}}}{\sum_{u \in c}\sum_{\alpha=1}^n{a_{u\alpha}}}.
\end{equation}
Since $u$ may select more than one object, we obtain $\langle p(o,u)\rangle$ by averaging $p(o,u)$ over all his/her collected objects. The dependence of $\langle p(o,u)\rangle$ on $k_u^{(o)}$ in both data sets is presented in fig. \ref{Fig:select_rate}.

The negative correlation in fig. \ref{Fig:select_rate} indicates that the small degree users (i.e. users with small $k_u^{(o)}$) usually have large $\langle p(o,u)\rangle$ while those large degree users tend to have small $\langle p(o,u)\rangle$ in both datasets. That is to say, groups have more potential influence on small degree users when they select objects. Moreover, by comparing $\langle p(o,u)\rangle$ of Last.fm and Douban, we find that the former is much larger than the latter. It means that users in Last.fm are much more active in groups and more likely to be influenced by groups.

Based on the analysis above, we conclude that the information of social groups can compensate the sparse data in the user-object network. When a user didn't choose any object or chose very few objects, we can still use the group information to obtain the similarity between him/her and other users. This is important since it may solve the user cold-start problem in information filtering \cite{Adomavicius:2005}.

\section{Information filtering}
\label{Sec:method}
In recent years, the study of information filtering attracts attention of researchers from different fields including social and computer scientists, physicists, and interdisciplinary researchers \cite{Lv:RP:2012}. As we mentioned in the introduction, one of the biggest challenges in the information filtering is the user cold-start problem which is very important from the commercial point of view. Solving it not only increases the loyalty of new users, but also stimulates the inactive users to use the web site more frequently. Previous recommendation algorithms only take into account the user-object network, which makes the information of the new/inactive users very limited \cite{Zhou:PRE:2007,Zhou:PNAS:2010}. Therefore, it is necessary to include also users' membership information \cite{Adomavicius:2005}. Motivated by the empirical analysis above, we propose a social diffusion recommendation algorithm (short for SD algorithm) for improving the recommendation accuracy of small degree users, i.e., solving the user cold-start problem. The basic idea is that one can predict users' preferences of objects through their membership information though they collected very few objects in the history.

\begin{figure*}
\centering
\includegraphics[width=6cm,height=3.2cm]{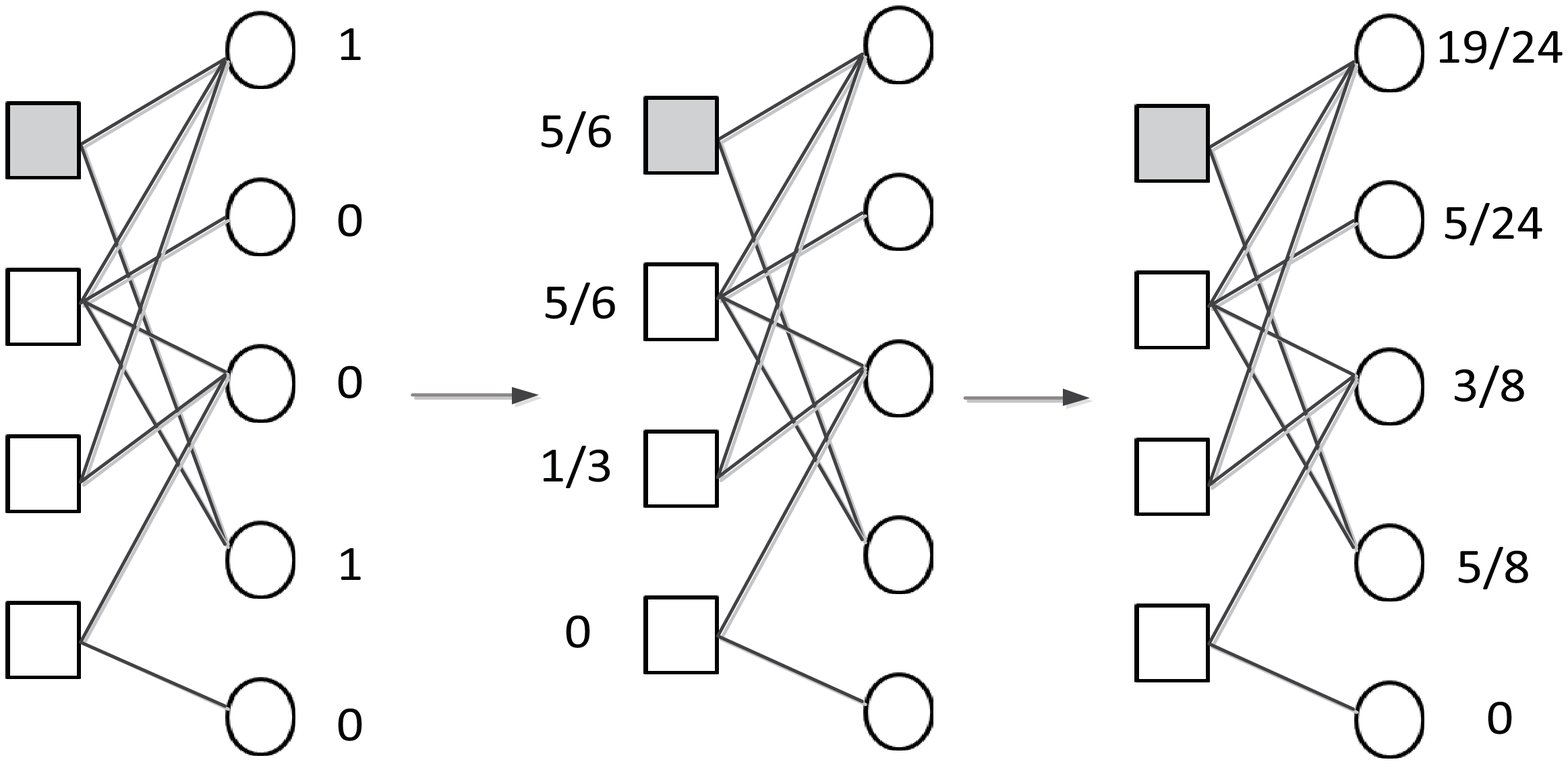}\hspace{1.5cm}
\includegraphics[width=7cm,height=3.2cm]{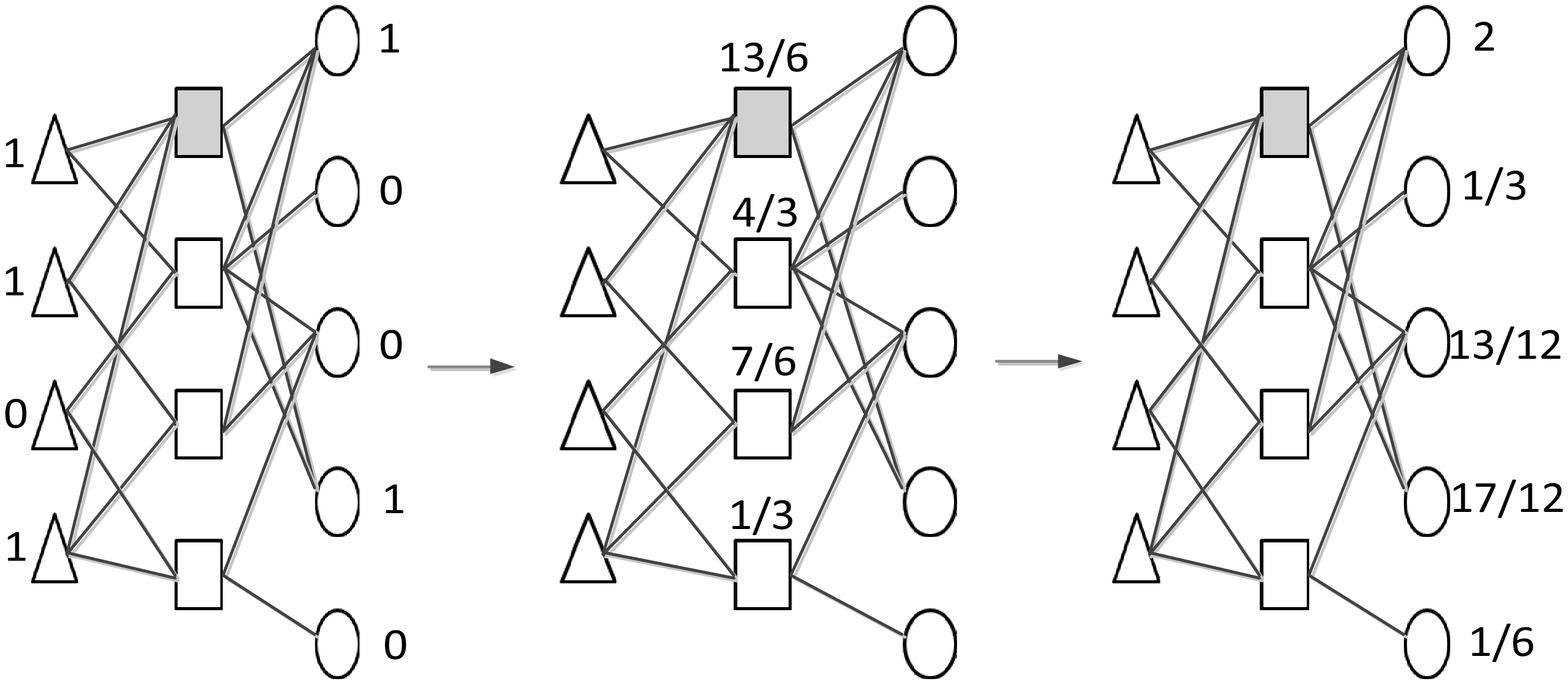}\\
\mbox{(a)Mass Diffusion\hspace{5cm}(b)Social Diffusion}\\
\caption{The illustration of the Mass Diffusion and the Social Diffusion processes. The gray square denotes the target user. Circles represent objects and triangles are the groups.}
\label{Fig:method}
\end{figure*}

The SD algorithm will be built on the mass diffusion process on both the user-object and user-group bipartite networks. In the original mass diffusion algorithm \cite{Zhou:PRE:2007}, given a target user $i$ to whom we will recommend objects to, each of $i$'s collected object is assigned with one unit of resource and they are equally distributed to all the neighboring users who have selected this object. If user $j$ is one of these users, the resource he/she received from $o$ will be $1/k_{o}$ where $k_o$ is degree of $o$ (namely the number of users who collected $o$). The final resource $j$ received is the sum over all $i$'s collected objects:
\begin{equation}
f_{j}^{(o)} = \sum_{o \in \Gamma^{(o)}_i\cap \Gamma_j^{(o)}}{\frac{a_{io}}{k_{o}}},
\end{equation}
where $\Gamma^{(o)}_i$ and $\Gamma^{(o)}_j$ are $i$ and $j$'s collected object sets, respectively. We can assign resource to the group mates in the same way in our SD method. Suppose user $j$ has joined a same group $c$ as the target user $i$, $j$ will receive $1/k_c$ resource from $c$. The final resource $j$ received is the sum over all $i$'s joined groups:
\begin{equation}
f_{j}^{(c)} = \sum_{c \in \Gamma_i^{(c)}\cap \Gamma_j^{(c)}}{\frac{b_{ic}}{k_{c}}},
\end{equation}
where $\Gamma^{(c)}_i$ and $\Gamma^{(c)}_j$ are $i$ and $j$'s joined group sets, respectively. Since $i$'s neighbors from both group and object point of view are similar users to $i$, we combine these two kinds of resource as $f_j=f_j^{(o)}+f_j^{(c)}$. Finally, we let each user distribute their resource $f_j$ equally to the neighboring objects. The final resource object $o$ obtained is
\begin{equation}
f_o=\sum_{\scriptsize \mbox{$\begin{array}{c}
o \in \Gamma^{(o)}_j\end{array}$}}{\frac{f_j}{k_j^{(o)}}}
\end{equation}
where $k_j^{(o)}$ is the number of objects $j$ collected. The final resources of all objects will be sorted in descending order and the objects with most resources will be recommended. The SD process is illustrated in fig. \ref{Fig:method}(b).

Some other well-known recommendation algorithms based on only user-object bipartite network are selected to compare with our method. The first one is the original mass diffusion algorithm \cite{Zhou:PRE:2007} (short for MD). It can be expressed by the matrix form $\overrightarrow{f}' = W\overrightarrow{f}$, where $\overrightarrow{f}$ is the initial resource vector on objects and the element $w_{\alpha\beta}$ of $W$ is the resource that object $\alpha$ received from object $\beta$. The transition matrix $W$ can be computed by
\begin{equation}
w_{\alpha\beta} = \frac{1}{k_{\beta}}\sum_{i=1}^m{\frac{a_{i\alpha}a_{i\beta}}{k_{i}}}.
\end{equation}
The final resource vector $\overrightarrow{f}'$ will be sorted in the descending order and those objects with
most resources will be recommended. See figure \ref{Fig:method}(a) for illustration.

The second one is the hybrid method combining the mass diffusion process \cite{Zhou:PRE:2007} and the heat conduction process \cite{Zhou:PNAS:2010} (short for HDH). Different from the pure mass diffusion algorithm, the transition matrix $W$ in the hybrid method is calculated by
\begin{equation}
w_{\alpha\beta} = \frac{1}{k^{1-\lambda}_{\alpha}k^{\lambda}_{\beta}}\sum_{i=1}^m{\frac{a_{i\alpha}a_{i\beta}}{k_{i}}},
\end{equation}
where $\lambda$ is a tunable parameter. If $\lambda=0$, it degenerates to the pure heat conduction algorithm
\cite{Zhang:PRL:2007}. If $\lambda=1$, it gives the mass diffusion algorithm \cite{Zhou:PRE:2007}.

Two kinds of collaborative filtering (CF) methods are also considered: the user-based CF (short for \emph{UCF}) and the item-based CF (short for \emph{ICF}). In UCF, the basic assumption is that similar users usually collect the same items. Accordingly, the recommendation score of object $\alpha$ for target user $i$ is
\begin{equation}
p_{i\alpha} = \sum_{j=1}^m{s_{ij}a_{j\alpha}},
\end{equation}
where $s_{ij}$ is the similarity between user $i$ and user $j$. Actually, the measure of similarities between two nodes in a network
is subject to the definition~\cite{Zeng:EPL2012}. In this paper, we apply the \emph{Salton} index to measure the similarity between users~\cite{Salton1983}:
\begin{equation}
s_{ij} = \sum_{\alpha=1}^n{\frac{a_{i\alpha}a_{j\alpha}}{\sqrt{k_ik_j}}}.
\end{equation}
Different from the UCF, the basic assumption of ICF is that a user is usually interested in the object similar to the objects already collected by him/her. The recommendation scores of $\alpha$ for target user $i$ is
\begin{equation}
p_{i\alpha}=\sum_{\beta=1}^n{s_{\alpha\beta}a_{i\beta}},
\end{equation}
where $s_{\alpha\beta}$ is the similarity between $\alpha$ and $\beta$ and computed also by the Salton index,
\begin{equation}
s_{\alpha\beta} = \sum_{i=1}^m{\frac{a_{i\alpha}a_{i\beta}}{\sqrt{k_{\alpha}k_{\beta}}}}.
\end{equation}

To test the recommendation accuracy of different algorithms, the links in the user-object bipartite network are randomly
divided into two parts: the training set ($E^T$) and the probe set ($E^P$). The training set contains $80\%$ of the original data and the recommendation algorithm runs on it. The left $20\%$ links forms the probe set which will be used to test the performance of the recommendation results. Note that the node sets (i.e. user sets and items sets) are equal in the training and probe sets.

We use the Ranking Score metric (RS) to test the accuracy of algorithms. As discussed above, each recommendation algorithm can provide each user an ordered list of his/her uncollected objects. For a user $i$, if the object $\alpha$ is in the probe set, we measure the position of this object $\alpha$ in the order list. For example, if there are $1000$ uncollected objects for $i$ and $\alpha$ is the $30$th from the top in the order list, we say the position of $\alpha$ is the top $30/1000$, and the ranking score $R_{i\alpha}=0.03$. A good algorithm is expected to give a small ranking score.

\begin{figure}[htbp]
\centering
\includegraphics[width=9cm,height=8cm]{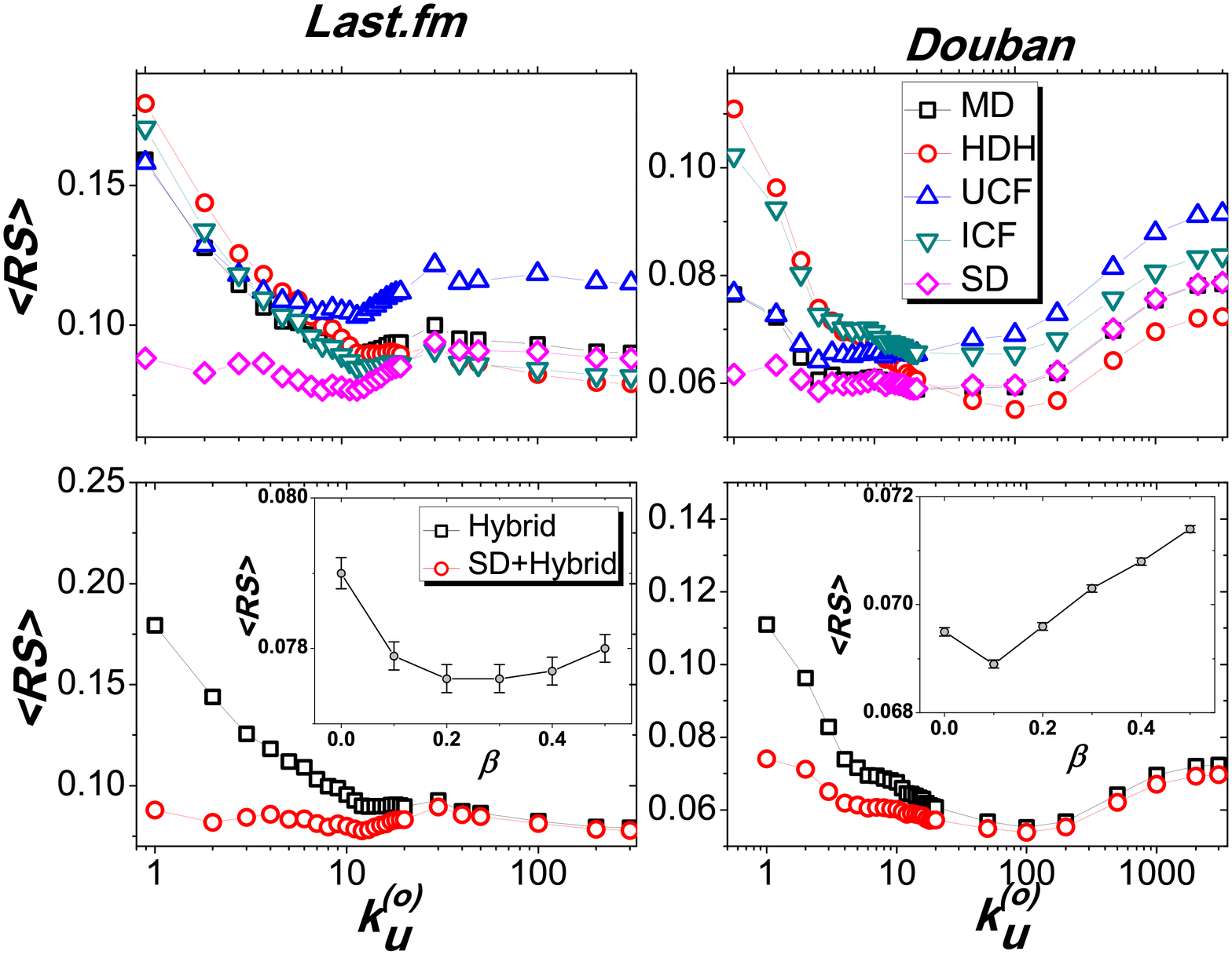}
\caption{Dependence of mean cumulative ranking score $\langle RS \rangle$ on the user degree $k_u^{(o)}$. Given a user degree $k_u^{(o)}$, $\langle RS \rangle$ is obtained by averaging the ranking score over all the users whose degrees are no larger than $k_u^{(o)}$. The optimal $\lambda$ of HDH is 0.3 for Douban and 0.4 for Last.fm, respectively. The insets are the dependence of overall $\langle RS\rangle$ on the parameter $\beta$ in SD+HDH. The error bars are the standard deviation of $\langle RS\rangle$ in different division of probe sets and training sets. In the bottom sub-figures, $\beta$ is set as the optimal value according to the results in the insets.}
\label{Fig:user_rs}
\end{figure}

The top sub-figures of fig. \ref{Fig:user_rs} give the relationship between the user degree $k_u^{(o)}$ and the mean cumulative ranking score $\langle RS \rangle$. Given a user degree $k_u^{(o)}$, the mean cumulative ranking score $\langle RS \rangle$ is obtained by averaging over all the users whose degrees are no larger than $k_u^{(o)}$. In fig. 4, it shows that the hybrid method gives the best mean ranking score over all users (see the $\langle RS \rangle$ when user degree is maximum). Compared to the MD method, the SD method works better in overall ranking score. Among all these methods, the UCF gives the worst overall ranking score.

In recommender systems, it is usually difficult to recommend objects to the users who have collected a few objects. For the small degree users, the accuracy of the hybrid method (HDH) is not so good as the other methods in both datasets. The UCF is better than the ICF when recommending objects to those small-degree users. The performance of the MD method is similar to the the UCF method. Among all methods considered, the SD method achieves the lowest ranking score when recommending objects for small-degree users, which indicates that the SD method is very effective in solving the user cold-start problem. Actually, considering the social grouping in recommendation will bring much value information for the small degree users. Compared to small degree users, large degree users have less common tastes with their group mates. As shown in fig. 2, $\langle p(o,u)\rangle$ keeps decreasing with $k_u^{(o)}$. Therefore, the information of groups should be considered less when recommending objects for large degree users.

Based on the results above, we conclude that the use of recommendation algorithms should be personalized. In other words, it is better to apply different recommendation algorithms for different users. For instance, we can use the social diffuse method to generate recommendations for the small degree users and use the hybrid method for large degree users. In this way, we can effectively improve the recommendation accuracy for the small degree users. At the same time, the recommendation accuracy for the large degree users can be well preserved. Here, we propose a personalized combination of SD and HDH methods. Denote $h_{i\alpha}$ and $g_{i\alpha}$ as the resource item $\alpha$ received from user $i$ by HDH and SD methods, respectively. The final recommendation score of item $\alpha$ to user $i$ can be expressed as: $f_{i\alpha}= \lambda_i h_{i\alpha}+(1-\lambda_i)g_{i\alpha}$, where $\lambda_i=(\frac{k_i^{(o)}}{max(k^{(o)}})^{\beta}$ and $k_i^{(o)}$ is the degree of user $i$. When $\beta=0$, all the users are using HDH. When $\beta$ is infinitely large, all the users are using SD. As $\beta$ increases, the recommendation method changes smoothly from HDH to SD, and the small degree users change faster than large degree users. As such, SD will have more weight in small degree users' recommendation. This combination is denoted as SD+HDH. The result is presented in the bottom sub-figures of fig. \ref{Fig:user_rs}. From the insets, one can see that the overall $\langle RS\rangle$ is improved by adjusting $\beta$ and there is an optimal $\beta$. The small error bars indicate that the optimal $\beta$ is stable in different divisions of probe set and training set. Therefore, in practical use, the future optimal $\beta$ can be estimated by testing the historical data. For the small degree users, the recommendation accuracy can be significantly improved by the SD+HDH method. We recall that users in the Last.fm are more active in communities. Therefore, the new method achieves a better accuracy improvement in Last.fm than in Douban dataset.

\section{Conclusion}
In summary, we investigate the online system coupled with user-object and user-group bipartite networks. Our results show that users may join in many groups though they have collected a few objects. Based on the \emph{Aspect Model}~\cite{Hofmann:2003:SIGIR}, we find that the the group mates of the small degree users share very similar tastes with them (i.e., their selected objects are similar). We further propose a recommendation method which
takes into account the information of users' membership (i.e.
the group that users joined). Our method
can largely improve recommendation accuracy for the small degree
users. However, this social diffusion method doesn't work well for the large degree
users. By combining the new method and the hybrid
method in~\cite{Zhou:PNAS:2010}, we achieve a higher recommendation
accuracy than the original hybrid method itself, especially
on the small-degree users. The user cold-start problem is
effectively solved by our method. Finally, we remark that
this work highlights that different users should be assigned
with their own suitable recommendation methods, which
may lead to a significant improvement of the recommendation
performance in the future.

\section*{Acknowledgments}
This work is supported by the opening foundation of Institute of Information Economy in Hangzhou Normal University (Grant No. PD12001003002002) and the Sichuan Provincial Science and Technology Department (Grant No. 2012FZ0120). W.Z. acknowledges the support from Sino-Swiss Science and Technology Cooperation Program (EG57-092011). A.Z. acknowledges the support from China scholarship Council.


\end{document}